\documentclass[12pt,reqno]{amsart}

\usepackage{amsthm,amssymb,amstext,amscd,euscript,mathrsfs,amsfonts,amsbsy,amsxtra,latexsym,amsmath}
\usepackage{fullpage}
\usepackage[english]{babel}
\usepackage[latin1]{inputenc}
\usepackage{verbatim}
\usepackage{graphicx} 
\usepackage[all]{xy} 
\usepackage{enumitem}
\usepackage{bbm}
\usepackage{marvosym}
\numberwithin{figure}{section} 
\allowdisplaybreaks

\newcommand{\field}[1]{\mathbb{#1}} 

\newcommand{\Z}{\field{Z}} 

\newcommand{\C}{\field{C}}

\numberwithin{equation}{section}

\numberwithin{theorem}{section}

\theoremstyle{definition}

\newtheorem{remark}{Remark}[section]

\newcommand{\bea}{\begin{eqnarray}} 
\newcommand{\eea}{\end{eqnarray}} 
\newcommand{\be}{\begin{equation}} 
\newcommand{\ee}{\end{equation}} 
\newcommand{\benn}{\begin{equation*}} 
\newcommand{\eenn}{\end{equation*}}

\title[short title]{On the Sen limit squared}
\author{James Fullwood}
\address{School of Mathematical Sciences\\Shanghai Jiao Tong University\\ 800 Dongchuan Road, Shanghai, China}
\email{fullwood@sjtu.edu.cn}
\author{Dongxu Wang}
\address{Department of Mathematics\\Dongbei University of Economics and Finance\\ 217 Jianshan St, Shahekou District, Dalian, China}
\email{dxwang@dufe.edu.cn}
\begin{document}

\maketitle
%%\begin{center}
%%\emph{Dedicated to Henry Laufer on his 70th birthday}
%%\end{center}

\begin{abstract}
We introduce a class of F-theory vacua which may be viewed as a specialization of the so-called $E_6$ fibration, and construct a weak coupling limit associated with such vacua which we view as the `square' of the Sen limit. We find that while Sen's limit is naturally viewed as an orientifold theory, the universal tadpole relation which equates the D3 charge between the associated F-theory compactification and the limit we construct suggests that perhaps the limiting theory is in fact an oriented theory compactified on the base of the F-theory elliptic fibration.  
\end{abstract}

\section{Introduction}
F-theory compactified on an elliptic Calabi-Yau $(n+1)$-fold $Y\to X$ is essentially a type-IIB compactification on the base $X$ with varying axio-dilaton \cite{VFT}. The $\text{SL}_2(\Z)$-invariant value of the axio-dilaton at a point $p\in X$ is then interpreted as the complex structure modulus of the elliptic fiber over $p$. If the Calabi-Yau $(n+1)$-fold $Y$ is a Weierstrass fibration, i.e., if it globally given by an equation of the form
\[
y^2z=x^3+fxz^2+gz^3,
\] 
Sen constructed in a systematic way how to identify F-theory compactified on $Y$ with a weakly-coupled orientifold theory compactified on a Calabi-Yau $n$-fold which is a two-sheeted cover of $X$ \cite{Sen}. The identification was established via an explicit deformation of the total space $Y$ of the elliptic $(n+1)$-fold to a degenerate Calabi-Yau in which all the fibers are singular, which signals weak coupling everywhere on the base. This systematic procedure via deformation of a Weierstrass equation for establishing a duality between F-theory and orientifold type-IIB theories is then referred to as the \emph{Sen limit}.

Since every elliptic fibration is birational to a fibration in Weierstrass form, for a while Sen's limit was viewed as being applicable to a general F-theory compactification. But as pointed out in \cite{AE1}\cite{AE2}, birational transformations of elliptic fibrations don't preserve singular fibers, and since the singular fibers of an elliptic fibration play a crucial role in the physics of F-theory, one is not investigating the general case by considering fibrations in Weierstrass form whose total space is smooth. Moreover, the Weierstrass form of a smooth elliptic fibration admits a total space which is in general singular, so in dealing with Weierstrass models of elliptic fibrations not initially given by a global Weierstrass equation, one must either resolve singularities or take up the issue of defining F-theory on singular elliptic fibrations \cite{EYSU5}\cite{TFSt}\cite{CSSF}\cite{TGPMil}. 

In light of this, in \cite{AE2} Aluffi and Esole initiated a program of moving beyond Weierstrass models in F-theory, where they considered families of smooth elliptic $n$-folds not given by global Weierstrass equations, and constructed orientifold limits associated with such fibrations by generalizing the method first employed by Sen. Further investigations into non-Weierstrass fibrations and weak coupling limits involving the Tate form of Weierstrass fibrations were explored in \cite{DWHBUV}\cite{ESTF}\cite{EFY}\cite{CCG}\cite{GTSGUT}. A comparison of the D3 charge in F-theory and its limit poses consistency conditions which take the form of an identity relating the Euler characteristic of the elliptic $(n+1)$-fold associated with the F-theory compactification and the Euler characteristics of D-branes which arise in the limit. For a general deformation of an elliptic Calabi-Yau, such consistency conditions -- referred to as \emph{tadpole relations} -- will not hold, so constructing a consistent limit is a non-trivial affair. Moreover, what is interesting from a purely mathematical perspective is that when the tadpole relations \emph{do} hold between an F-theory compactification and its limit, in all known examples the associated identity between Euler characteristics turns out to be the dimension-zero component of a much more general identity which holds at the level of Chern classes, and moreover, the identities hold without any dimensional constraints on the base of the fibration and without any Calabi-Yau hypothesis on the total space \cite{AE1}\cite{AE2}\cite{EFY}\cite{EJY}. Such Chern class identities were then referred to as \emph{universal tadpole relations}, and a purely mathematical explanation for the existence of such identities was given in \cite{FTR2}.
\\

In this note, we introduce a class of F-theory vacua that, while not given by a global Weierstrass equation, its discriminant is given by an equation which may be smoothly deformed to the \emph{square} of the discriminant of a smooth Weierstrass fibration. In particular, we construct a weak coupling limit associated with such fibrations in such a way that the limiting discriminant is the square of the limiting discriminant in Sen's limit, and as such, the the branes arising in the limit are all doubles of the branes arising in Sen's limit. The main difference from Sen's limit however, is we find that the universal tadpole relation associated with the limit holds only once a factor of two that appears when viewing charges from the perspective of an orientifold theory is canceled. So while an orientifold interpretation of the limit may exist, the form of the universal tadpole relation suggests to us that the limit we construct may exist more naturally as an oriented theory on the base of the orientifold projection, i.e., on the `square' of the orientifold. 
\\

\noindent \emph{Acknowledgements}. JF would like to thank Mboyo Esole and Matt Young for useful discussions.

\section{Sen's limit}\label{SL}
The $F$-theory vacua considered by Sen in constructing his limit correspond precisely to smooth Weierstrass models, i.e., elliptic fibrations associated with the Weierstrass equation
\begin{equation}\label{we}
y^2z=x^3+fxz^2+gz^3.
\end{equation}
For equation (\ref{we}) to define an elliptic fibration, let $X$ be a smooth compact Fano $n$-fold over $\C$, so that its anti-canonical bundle $\mathscr{O}(-K_X)$ is ample. We then consider the vector bundle\footnote{A super-script on a line bundle such as $\mathscr{L}^k$ will always be used to denote the $k$th tensor power of the line bundle for any positive integer $k$.}
\[
\mathscr{E}=\mathscr{O}\oplus \mathscr{O}(-K_X)^2\oplus \mathscr{O}(-K_X)^3,
\]
 and denote by $\pi:\mathbb{P}(\mathscr{E})\to X$ the projective bundle of \emph{lines} in $\mathscr{E}$. The tautological bundle on $\mathbb{P}(\mathscr{E})$ will then be denoted by $\mathscr{O}(-1)$. We then associate with equation (\ref{we}) a section of $\mathscr{O}(3)\otimes \pi^*\mathscr{O}(-K_X)^6$ by taking $x$ to be a section of $\mathscr{O}(1)\otimes \pi^*\mathscr{O}(-K_X)^2$, $y$ to be a section of $\mathscr{O}(1)\otimes \pi^*\mathscr{O}(-K_X)^3$, $z$ to be a section of $\mathscr{O}(1)$, $f$ to be a section of $\pi^*\mathscr{O}(-K_X)^4$ and $g$ to be a section of $\pi^*\mathscr{O}(-K_X)^6$. With these prescriptions, equation (\ref{we}) then defines a codimension 1 embedding $W\hookrightarrow \mathbb{P}(\mathscr{E})$, and composing the embedding with the bundle projection $\pi:\mathbb{P}(\mathscr{E})\to X$ endows $Y$ with the structure of an elliptic $(n+1)$-fold $Y\to X$, where the fiber over a point $p$ in $X$ is given by
 \[
 W_p:(y^2z=x^3+f(p)xz^2+g(p)z^3)\subset \mathbb{P}^2.
 \]   
 For $W$ to be smooth we require that the hypersurfaces
 \[
 F:(f=0)\subset X \quad \text{and} \quad G:(g=0)\subset X
 \]
 are both smooth and intersect transversally. The singular fibers of $W\to X$ then lie over the discriminant hypersurface
 \[
 \Delta_W:(4f^3+27g^2=0)\subset X.
 \] 
 Over a general point of $\Delta_W$, the fiber of $W\to X$ will be a nodal cubic, and over the codimension 2 smooth complete intersection
 \[
 C:(f=g=0)\subset X
 \]
 the fibers enhance to cuspidal cubics. From B.5.8 in \cite{IT} along with the adjunction formula 
 \[
 c(W)=\frac{(1+H)(1+H-2K_X)(1+H-3K_X)(3H-6K_X)}{1+3H-6K_X}\pi^*c(X),
 \]
 where $H$ denotes the first Chern class of the bundle $\mathscr{O}(1)\to \mathbb{P}(\mathscr{E})$. It immediately follows that the first Chern class of $W$ is zero, so that the total space of the fibration is an anti-canonical hypersurface in $\mathbb{P}(\mathscr{E})$. 
 
From the type-IIB viewpoint, the axio-dilaton field on $X$ is given by
\[
\tau=C_{(0)}+i\frac{1}{g_s},
\] 
where $C_{(0)}$ is the axion RR-scalar field and $g_s$ is the string coupling constant. From the $F$-theory viewpoint the axio-dilaton $\tau$ is then identified with the complex structure modulus of the elliptic fibers of $W\to X$. A weak coupling limit then corresponds to deforming $W$ in such a way that the complex structure modulus approaches $i\infty$ everywhere over $X$. Sen's limit is then constructed by perturbing $f$ and $g$ in equation (\ref{we}) in terms of a complex deformation parameter $t$ in such a way that the central fiber of the associated family is a degenerate fibration in which all the fibers are singular (so that $\tau$ approaches $i\infty$). In particular, the family corresponding to Sen's limit is given by 
 \[
\mathscr{W}:(y^2z=x^3+ (-3h^2+t\eta)xz^2+(-2h^3+th\eta+t^2\psi)z^3)\subset \mathbb{P}(\mathscr{E})\times D,
\] 
where $D$ denotes an open disk centered about $0\in \C$, and $h$, $\eta$ and $\psi$ are general sections of $\pi^*\mathscr{O}(-K_X)^2$, $\pi^*\mathscr{O}(-K_X)^4$ and $\pi^*\mathscr{O}(-K_X)^6$ respectively. The central fiber is then given by
 \[
W_0:(y^2z=x^2+(-3h^2)xz^2+(-2h^3)z^3)\subset \mathbb{P}(\mathscr{E}),
\] 
whose fibers are nodal for $h\neq 0$ and cuspidal for $h=0$. The auxiliary $n$-fold corresponding to the orientifold theory is then given by
\[
Z:(\zeta^2=h)\subset \mathscr{O}(-K_X),
\]
where $\zeta$ is a general section $\mathscr{O}(-K_X)$. It then follows that $Z$ is a double cover of $X$ ramified over $O:(h=0)\subset X$, which we identify with the orientifold plane. The tangent bundle of the total space of the anti-canonical bundle $p:\mathscr{O}(-K_X)\to X$ fits into the exact sequence 
\[
0\to p^*\mathscr{O}(-K_X)\to T\mathscr{O}(-K_X)\to p^*TX\to 0,
\]  
thus by adjunction we have
\[
c(Z)=p^*\left(\frac{c(\mathscr{O}(-K_X))c(TX)}{c(\mathscr{O}(-2K_X))}\right)\cap [Z].
\]
It immediately follows that $c_1(Z)=0$, so $Z$ is in fact a Calabi-Yau $n$-fold.

To locate the D-branes associated with the limit, we take the flat limit as $t\to 0$ of the corresponding family of discriminants 
\[
\mathscr{D}_W:(4(-3h^2+t\eta)^3+27(-2h^3+th\eta+t^2\psi)^2=0)\subset X\times D, 
\] 
which yields
\[
\Delta_{W_0}:(h^2(\eta^2+12h\psi)=0)\subset X.
\]
We then pullback the flat limit $\Delta_0$ of $\mathscr{D}_W$ as $t\to 0$  to $Z$ to obtain the D-brane spectrum of the orientifold limit, which is given by
\[
\widetilde{\Delta_{W_0}}:(\zeta^4(\eta^2+12\zeta^2\psi)=0)\subset Z.
\] 
The D-brane given by
\[
D:(\eta^2+12\zeta^2\psi=0)\subset Z
\]
has singularities reminiscent of those of a Whitney umbrella, and as such was referred to as a `Whitney brane' in \cite{AE1}\cite{CDM}. Comparing the D3 tadpole condition between $F$-theory and type-IIB then predicts that $2\chi(W)$ coincides with $4\chi(O)+\chi(D)$ ($\chi$ denotes topological Euler characteristic with compact support), but it was shown in the literature that for such a relation to hold a certain contribution is needed coming from the singularities of $D$ \cite{CDM}. In particular, it was shown that such tadpole relations hold precisely when the pinch-point singularities of $D$ are subtracted from the charge of a small resolution $\overline{D}$ of $D$, thus yielding the tadpole relation
\begin{equation}\label{tr}
2\chi(W)=4\chi(O)+\chi(\overline{D})-\chi(S),
\end{equation}
where $S$ is the pinch locus of $D$, namely
\[
S:(\zeta=\eta=\psi=0)\subset Z.
\]
It was then shown by Aluffi and Esole that the tadpole relation (\ref{tr}) is in fact a consequence of a much more general relation at the level of Chern classes, which they referred to as a \emph{universal tadpole relation} \cite{AE2}. More precisely, let $\psi:W\to X$ denote the projection of the elliptic fibration corresponding to the F-theory vacua, then the tadpole relation (\ref{tr}) is just the dimension-zero component of the universal tadpole relation
\begin{equation}\label{utr}
2\psi_*c(W)=4c(O)+c^{(\infty)}(D),
\end{equation}
where $c^{(\infty)}(D)$ denotes a certain characteristic class of the singular variety $D$ whose component of dimension zero coincides with $\chi(\overline{D})-\chi(S)$ as appearing on the RHS of (\ref{tr}) (see \S5 of \cite{AE1} for a the precise definition of $c^{(\infty)}(D)$). We can also write formula \eqref{utr} is in terms of the components of the limiting discriminant $\Delta_{W_0}$. In particular, denote by $\underline{D}$ the singular component of $\Delta_{W_0}$ given by
\[
\underline{D}:(\eta^2+12h\psi=0)\subset X,
\]
and denote its singular locus by $\underline{S}$, which is a smooth codimenison 3 complete intersection given by
\[
\underline{S}:(h=\eta=\psi=0)\subset X.
\]
Then it turns out that $c^{(\infty)}(D)=2(c_{\text{SM}}(\underline{D})-c(\underline{S}))$ (again see \cite{AE1} \S5 for details), thus \eqref{utr} may be rewritten as
\begin{equation}\label{utr2}
2\psi_*c(W)=4c(O)+2(c_{\text{SM}}(\underline{D})-c(\underline{S})),
\end{equation}
where $c_{\text{SM}}(\underline{D})$ denotes the Chern-Schwartz-MacPherson class of $\underline{D}$ \footnote{A nice introduction to Chern-Schwartz-MacPherson classes is given in \cite{CSVA}.}. This form \eqref{utr2} of the universal tadpole relation for Sen's limit will be of use when we make a connection with the limit constructed in \S\ref{SRL}.

\section{A specialization of the $E_6$ fibration}\label{NC}
Let $X$ be a smooth compact Fano $n$-fold over $\C$ as in \S\ref{SL}. We now introduce a class of $F$-theory vacua over $X$ from which we will construct a weak coupling limit which may be viewed as the square of Sen's limit. For this, we use the equation
\begin{equation}\label{ne}
x^3+y^3+eyz^2+fxz^2+gz^3=0.
\end{equation}
We then consider the vector bundle
\[
\mathscr{E}=\mathscr{O}\oplus \mathscr{O}(-K_X)\oplus \mathscr{O}(-K_X),
\] 
and denote by $\pi:\mathbb{P}(\mathscr{E})\to X$ the projective bundle of lines in $\mathscr{E}$, with tautological bundle $\mathscr{O}(-1)\to \mathbb{P}(\mathscr{E})$. For equation (\ref{ne}) to define the zero-locus of a well-defined section of a line bundle on $\mathbb{P}(\mathscr{E})$, we take $x$ and $y$ both to be sections of $\mathscr{O}(1)\otimes \pi^*\mathscr{O}(-K_X)$, $z$ to be a section of $\mathscr{O}(1)$, both $e$ and $f$ to be a section of $\pi^*\mathscr{O}(-K_X)^2$, and $g$ to be a section of $\pi^*\mathscr{O}(-K_X)^3$. With these prescriptions, the LHS of equation (\ref{ne}) yields a well-defined section of $\mathscr{O}(3)\otimes \pi^*\mathscr{O}(-K_X)^3$, whose zero locus defines a codimension 1 embedding $Y\hookrightarrow \mathbb{P}(\mathscr{E})$. Composing the embedding $Y\hookrightarrow \mathbb{P}(\mathscr{E})$ with the bundle projection $\pi:\mathbb{P}(\mathscr{E})\to X$ yields the projection $\varphi:Y\to X$, which endows the total space $Y$ with the structure of an elliptic fibration. The fiber over a point $p\in X$ is then given by
\[
Y_p:(x^3+y^3+e(p)yz^2+f(p)xz^2+g(p)z^3=0)\subset \mathbb{P}^2.
\] 
The Chern class of $Y$ is given by
\[
c(Y)=\frac{(1+H)(1+H-K_X)^2(3H-3K_X)}{1+3H-3K_X}\pi^*c(X),
\]
where again $H$ denotes the first Chern class of $\mathscr{O}(1)\to \mathbb{P}(\mathscr{E})$. It immediately follows that $c_1(Y)=0$, so that $Y$ is an anti-canonical hypersurface in $\mathbb{P}(\mathscr{E})$.

The connection with Sen's limit comes from the fact that by setting $e=0$, the discriminant of $\varphi:Y\to X$ becomes the square of the discriminant of Weierstrass fibrations (which will be crucial in the next section, when we construct a weak coupling limit). In particular, we have that the Weierstrass form for $Y$ is given by
\[
y^2z+x^3+Fxz^2+Gz^3=y^2z+x^3+(-3ef)xz^2+\left(f^3+\frac{27}{4}g^2\right)z^3=0,
\]
so that $F=-3ef$ and $G=e^3+f^3+\frac{27}{4}g^2$. The discriminant of $\varphi:Y\to X$ is then given by
\[
\Delta_Y:(4F^3+27G^2=0)\subset X.
\]
We note that when $e=0$, $\Delta_Y$ takes the form $(4f^3+27g^2)^2$, which is the square of the discriminant of Weierstrass fibrations $\Delta_W$. The $j$-invariant viewed as a function on $X\setminus \Delta_Y$ then takes the form
\[
j=1728\frac{e^3f^3}{4e^3f^3-(e^3+f^3+\frac{27}{4}g^2)^2}.
\]
As in the case of Weierstrass fibrations, in order to ensure that the total space of the fibration $Y$ is smooth we assume that the hypersurfaces 
 \[
 E:(e=0)\subset X \quad \text{and} \quad F:(f=0)\subset X \quad \text{and} \quad G:(g=0)\subset X
 \]
are all smooth and intersect each other transversally. The fibers over a general point of $\Delta_Y$ are nodal cubics which enhance to cuspidal cubics over $e=0$, while over the smooth codimension 3 complete intersection
 \[
 C:(e=f=g=0)\subset X
 \]
the fibers enhance to a bouquet of three 2-spheres intersecting at a point. 
 
 \begin{remark}\label{r1}
 The fibration $\varphi:Y\to X$ may be seen as a special case of the so-called $E_6$ fibration which was studied in detail in \cite{AE2}, which is given by
 \[
 Y_{E_6}:(x^3+y^3= dxyz+exz^2+fyz^2 + gz^3)\subset \mathbb{P}(\mathscr{E}).
\]
As such, we see that the fibration $\varphi:Y\to X$ corresponds to setting $d=0$ in the equation for $Y_{E_6}$. Moreover, since $Y_{E_6}$ may also be realized as the zero-locus of a section of $\mathscr{O}(3)\otimes \pi^*\mathscr{O}(-K_X)^3$, $Y_{E_6}$ has the same divisor class as $Y$ in $\mathbb{P}(\mathscr{E})$, thus they share the same Chern classes, Euler characteristic, Chern numbers etc. This is interesting in light of the fact that the $E_6$ fibration admits six topologically distinct singular fibers, while the fibration $\varphi:Y\to X$ admits only three.  
 \end{remark}
 
\section{Sen's limit squared}\label{SRL}
We now define a weak coupling limit associated with $\varphi:Y\to X$ by essentially deforming $e$ to 0, and perturbing $f$ and $g$ in the equation for $Y$ in precisely the same way as in Sen's limit. In particular, we take 
\[
e=t\epsilon, \quad f=-3\zeta^2+t\vartheta, \quad \text{and} \quad g=-2\zeta^3+t\zeta\vartheta+t^2\phi,
\]
where $\zeta$, $\vartheta$ and $\phi$ are all sections of the square roots of the line bundles for which $h$, $\eta$ and $\psi$ denoted sections of in Sen's limit respectively, and $t$ as before is a complex deformation parameter varying over a disk $D$ about the origin in $\C$. Such prescriptions then give rise to a family
\[
\mathscr{Y}:(x^3+y^3+t\epsilon yz^2+(-3\zeta^2+t\vartheta)xz^2+(-2\zeta^3+th\vartheta+t^2\phi)z^3=0)\subset \mathbb{P}(\mathscr{E})\times D,
\]
whose central fiber corresponds to the weak coupling limit, since in the limit $t\to 0$ the $j$-invariant approaches $\infty$ everywhere over the base. The limiting discriminant then takes the form
\begin{equation}\label{ld}
\Delta_{Y_0}:(\zeta^4(\vartheta^2+12\zeta\phi)^2=0)\subset X,
\end{equation}
which is essentially the same as the limiting discriminant in Sen's limit but with all components squared. 

At this point, one may define the auxiliary $n$-fold required for an orientifold compactification in precisely the same way as in Sen's limit, i.e., by defining a double cover of $X$ given by
 \[
 Z:(\zeta^2=h)\subset \mathscr{O}(-K_X),
 \] 
where $\zeta$ and $h$ now denote pull backs to $\mathscr{O}(-K_X)$ of sections of $\mathscr{O}(-K_X)$ and $\mathscr{O}(-K_X)^2$ respectively, so that the orientifold plane $O$ is given by $h=0$. But since the limiting discriminant $\Delta_{Y_0}$ appears with a factor of $\zeta^4$ rather than $h^2$ (as in the case of Sen's limit), its pullback to $Z$ doesn't change the form of its equation in any way. As such, it is perhaps more natural to view the limiting type-IIB theory as an oriented theory on the base $X$, rather than an orientifolded theory on $Z$. From this perspective,
the D-brane spectrum associated with the limit then consists of a stack of 4 branes supported on the smooth locus
\[
\mathcal{D}:(\zeta=0)\subset X,
\]
and a double-brane supported on
\[
D:(\vartheta^2+12\zeta\phi=0)\subset X.
\]
The D-branes supported on $D$ then admit singularities along the codimension 3 locus 
\[
S:(\vartheta=\zeta=\phi=0)\subset X.
\]

In the next section we derive the universal tadpole relation associated with this limit, which takes the form
\begin{equation}\label{utr3}
 \varphi_*c(Y)=4c(\mathcal{D})+2(c_{\text{SM}}(D)-c(S)).
\end{equation}
We point out that \eqref{utr3} is very similar to the universal tadpole relation \eqref{utr2} associated with Sen's limit, namely
\begin{equation}\label{utr4}
2\psi_*c(W)=4c(O)+2(c_{\text{SM}}(\underline{D})-c(\underline{S})).
\end{equation}
In particular, while the RHS of both equations are identical in form, the LHS of equation \eqref{utr3} differs from the LHS of \eqref{utr4} by a factor of 2, which perhaps further suggests that while the tadpole relation \eqref{utr4} corresponds to an orientifolded theory, the relation \eqref{utr3} in fact corresponds to an oriented one. 

 \section{Universal tadpole relation}\label{UTR}
We now derive the universal tadpole relation \eqref{utr3}. We first note that since the divisor class of the total space of the fibration $\varphi:Y\to X$ in $\mathbb{P}(\mathscr{E})$ is the same as the $E_6$ fibration mentioned in Remark~\ref{r1}, they share the same Chern classes. And since the pushforward to its base of the Chern class of an $E_6$ fibration was computed in Theorem~4.3 in \cite{AE2}, we already know $\varphi_*c(Y)$, which is given by
\begin{equation}\label{l0}
\varphi_*c(Y)=4c(G),
\end{equation}
where $G$ is the smooth hypersurface in $X$ given by $g=0$ (recall that $g$ is the coefficient of $z^3$ in the defining equation for $Y$). Moreover, viewing Lemma~4.4 in \cite{AE1} from a more general perspective, its conclusion says that if $V$ is a singular hypersurface in $X$ given by a general equation of the form
\[
V:(x^2+12yz=0)\subset X,
\]
and if we denote by $\mathcal{Y}$ and $\mathcal{S}$ the subvarieties of $X$ corresponding to the equations $y=0$ and $x=y=z=0$ respectively, then 
\begin{equation}\label{l1}
2c(\mathcal{Y})+c_{\text{SM}}(V)-c(\mathcal{S})=2c(\mathcal{G}),
\end{equation}
where $\mathcal{G}$ is a smooth hypersurface in $X$ whose divisor class is $3[\mathcal{Y}]$. As such, from the definition of $\mathcal{D}$, $D$ and $S$ in the previous section, in equation \eqref{l1} we may replace $\mathcal{Y}$ by $\mathcal{D}$, $V$ by $D$ and $\mathcal{S}$ by $S$ to arrive at the equation
\[
2c(\mathcal{D})+c_{\text{SM}}(D)-c(S)=2c(G),
\]
since $[G]=3[\mathcal{D}]$. Putting this equation together with equation \eqref{l0} then yields
\begin{equation}\label{l2}
\varphi_*c(Y)=4c(\mathcal{D})+2(c_{\text{SM}}(D)-c(S)),
\end{equation}
which we view as the universal tadpole relation associated with the limit constructed in the previous section \S\ref{SRL}. Note that the limiting discriminant associated with the limit is of the form
\[
\zeta^4(\vartheta^2+12\zeta\phi)^2=0,
\] 
so that the RHS of \eqref{l2} is precisely the sum of Chern classes of branes supported on the irreducible components of the limiting discriminant weighted with the appropriate multiplicities (together with the negative contribution coming from the singularities $S$ of $D$, as in Sen's limit). If the limit was then viewed as an oreintifold theory compactified on a double cover of $X$ given by $\zeta^2-h=0$, then the pullback to the orientifold of the limiting discriminant wouldn't change its form at all, thus we'd expect a similar relation to \eqref{l2} but with $\mathcal{D}$ replaced by the orientifold plane $O$ and a factor of 2 appearing on the LHS, as in the universal tadpole relation associated with Sen's limit, which we recall is given by
\[
2\psi_*c(W)=4c(O)+2(c_{\text{SM}}(\underline{D})-c(\underline{S})).
\] 

Perhaps there is a way to still view the limit as an orientifold theory by adding fluxes or more carefully investigating the nature of the D3 charge in the context at hand, but an oriented theory on the base seems more natural, at least from the form of equation \eqref{l2}.

\bibliographystyle{plain}
\bibliography{SSL2}

\begin{thebibliography}{10}

\bibitem{CSVA}
P.~Aluffi.
\newblock Characteristic classes of singular varieties.
\newblock In {\em Topics in cohomological studies of algebraic varieties},
  Trends Math., pages 1--32. Birkh\"auser, Basel, 2005.

\bibitem{AE1}
Paolo Aluffi and Mboyo Esole.
\newblock Chern class identities from tadpole matching in type {IIB} and
  {F}-theory.
\newblock {\em J. High Energy Phys.}, (3):032, 29, 2009.

\bibitem{AE2}
Paolo Aluffi and Mboyo Esole.
\newblock New orientifold weak coupling limits in {F}-theory.
\newblock {\em J. High Energy Phys.}, (2):020, i, 52, 2010.

\bibitem{TGPMil}
Philipp Arras, Antonella Grassi, and Timo Weigand.
\newblock Terminal singularities, {M}ilnor numbers, and matter in {F}-theory.
\newblock {\em J. Geom. Phys.}, 123:71--97, 2018.

\bibitem{CCG}
Sergio~L. Cacciatori, Andrea Cattaneo, and Bert van Geemen.
\newblock A new {CY} elliptic fibration and tadpole cancellation.
\newblock {\em J. High Energy Phys.}, (10):031, 20, 2011.

\bibitem{CDM}
Andr{\'e}s Collinucci, Frederik Denef, and Mboyo Esole.
\newblock D-brane deconstructions in {IIB} orientifolds.
\newblock {\em J. High Energy Phys.}, (2):005, 57, 2009.

\bibitem{CSSF}
Andr\'es Collinucci and Raffaele Savelli.
\newblock F-theory on singular spaces.
\newblock {\em J. High Energy Phys.}, (9):100, front matter+38, 2015.

\bibitem{DWHBUV}
Ron Donagi and Martijn Wijnholt.
\newblock Higgs bundles and {UV} completion in {$F$}-theory.
\newblock {\em Comm. Math. Phys.}, 326(2):287--327, 2014.

\bibitem{EFY}
Mboyo Esole, James Fullwood, and Shing-Tung Yau.
\newblock {$D_5$} elliptic fibrations: non-{K}odaira fibers and new orientifold
  limits of {F}-theory.
\newblock {\em Commun. Number Theory Phys.}, 9(3):583--642, 2015.

\bibitem{EJY}
Mboyo Esole, Monica Jinwoo~Kang, and S.-T. Yau.
\newblock A {N}ew {M}odel for {E}lliptic {F}ibrations with a {R}ank {O}ne
  {M}ordell-{W}eil {G}roup: I. {S}ingular {F}ibers and {S}emi-{S}table
  {D}egenerations.
\newblock {\em arXiv:1410.0003}, 2011.

\bibitem{ESTF}
Mboyo Esole and Raffaele Savelli.
\newblock Tate form and weak coupling limits in {F}-theory.
\newblock {\em J. High Energy Phys.}, (6):027, front matter+39, 2013.

\bibitem{EYSU5}
Mboyo Esole and Shing-Tung Yau.
\newblock Small resolutions of {$\rm SU(5)$}-models in {F}-theory.
\newblock {\em Adv. Theor. Math. Phys.}, 17(6):1195--1253, 2013.

\bibitem{FTR2}
James Fullwood.
\newblock On tadpole relations via {V}erdier specialization.
\newblock {\em J. Geom. Phys.}, 104:54--63, 2016.

\bibitem{IT}
William Fulton.
\newblock {\em Intersection theory}, volume~2 of {\em Ergebnisse der Mathematik
  und ihrer Grenzgebiete. 3. Folge.}
\newblock Springer-Verlag, Berlin, second edition, 1998.

\bibitem{GTSGUT}
Thomas~W. Grimm, Sven Krause, and Timo Weigand.
\newblock F-theory {GUT} vacua on compact {C}alabi-{Y}au fourfolds.
\newblock {\em J. High Energy Phys.}, (7):037, 33, 2010.

\bibitem{TFSt}
Craig Lawrie and Sakura Sch\"afer-Nameki.
\newblock The {T}ate form on steroids: resolution and higher codimension
  fibers.
\newblock {\em J. High Energy Phys.}, (4):061, front matter+65, 2013.

\bibitem{Sen}
Ashoke Sen.
\newblock Orientifold limit of {F}-theory vacua.
\newblock {\em Nuclear Phys. B Proc. Suppl.}, 68:92--98, 1998.
\newblock Strings '97 (Amsterdam, 1997).

\bibitem{VFT}
Cumrun Vafa.
\newblock Evidence for {$F$}-theory.
\newblock {\em Nuclear Phys. B}, 469(3):403--415, 1996.

\end{thebibliography}

\end{document}